%% file: main.tex
\documentclass[usenatbib]{mn2e}
\usepackage{amsmath}
\usepackage{natbib}
\usepackage{epsfig}
\usepackage{txfonts}
\usepackage{pict2e}
\usepackage{cleveref}
\usepackage{color}
\usepackage{xspace}

\usepackage{url}

\newcommand{\omb}{\Omega_{\rm b} h^2}
\newcommand{\omc}{\Omega_{\rm c} h^2}
\newcommand{\thetaMC}{100\,\theta_{\rm MC}}
\newcommand{\ns}{n_{\rm s}}
\newcommand{\As}{A_{\rm s}}
\newcommand{\logA}{\ln\left(10^{10} A_{\rm s}\right)}
\newcommand{\aEMs}{\alpha_{\rm EM, 0}}
\newcommand{\mes}{m_{\rm e, 0}}
\newcommand{\aEM}{\alpha_{\rm EM}}

\newcommand{\LCDM}{$\Lambda$CDM\xspace}
\newcommand{\ho}{H_0}
\newcommand{\planck}{{\it Planck}\xspace}

\input{Befehle}

\title[Updates on fundamental constants from the CMB]
{Updated fundamental constant constraints from Planck 2018 data and possible relations to the Hubble tension}

\author[L.~Hart and J.~Chluba]{
Luke Hart$^{1}$\thanks{Email: luke.hart@manchester.ac.uk} and Jens Chluba$^{1}$ 
\\
$^{1}$Jodrell Bank Centre for Astrophysics, Alan Turing Building, University of Manchester, Manchester M13 9PL \\}

\date{\vspace{-0mm}Accepted  --. Received --.}

\pubyear{2019}

\begin{document}
\label{firstpage}
\pagerange{\pageref{firstpage}--\pageref{lastpage}}

\maketitle

\begin{abstract}
We present updated constraints on the variation of the fine structure constant, $\aEM$, and effective electron rest mass, $\me$, during the cosmological recombination era. These two fundamental constants directly affect the ionization history at redshift $z\simeq 1100$ and thus modify the temperature and polarisation anisotropies of the cosmic microwave background (CMB) measured precisely with {\it Planck }. 
The constraints on $\aEM$ tighten slightly due to improved {\it Planck} 2018 polarisation data but otherwise remain similar to previous CMB analysis.
However, a comparison with the 2015 constraints reveals a mildly discordant behaviour for $\me$, which from CMB data alone is found below its local value. Adding baryon acoustic oscillation data brings $\me$ back to the fiducial value, $\me=(1.0078\pm0.0067)\,m_{\rm e,0}$, and also drives the Hubble parameter to $\ho=69.1\pm 1.2$ [in units of ${\rm km \, s^{-1} \, Mpc^{-1} }$]. Further adding supernova data yields $\me=(1.0190\pm0.0055)\,m_{\rm e,0}$ with $\ho=71.24\pm0.96$. 
We perform several comparative analyses using the latest cosmological recombination calculations to further understand the various effects. 
Our results indicate that a single-parameter extension allowing for a slightly increased value of $\me$ ($\simeq 3.5\sigma$ above $m_{\rm e,0}$) could play a role in the Hubble tension.
\end{abstract}


\section{Introduction}

In the last few decades, we have achieved unprecedented cosmological results with the CMB anisotropies through various missions \citep{wmap9results, Planck2013params}. The 2015 release of {\it Planck} gave us unparalleled precision on the temperature spectra and greatly improved polarisation data at small angular scales \citep{Planck2015params, Planck2015like}. With these modern day developments, we have opened a gateway to various extensions to the standard $\Lambda$CDM model. Several groups have studied cosmological limits on neutrino masses and the number of relativistic degrees of freedom, encoded in the $N_{\rm eff}$ parameter, leading to further discussion on the makeup of relativistic species in our universe \citep{PlanckNeutrino, BattyeNeutrinos}. Similarly, the advances in CMB anistropy data have allowed us to explore parameter space of Big Bang Nucleosynthesis as well as other physics beyond the standard model such as magnetic field heating \citep[e.g.,][]{Shaw2010PMF, Planck2016PMF}, non-standard recombination \citep[e.g.,][]{RubinoMartin2010, Farhang2013} and WIMP dark matter annihilation models \citep[e.g.,][]{Galli2009, Huetsi2009, Chluba2010a, Planck2015params}. 

One of the many key physical processes that we can study during the recombination epoch is the possible variations of fundamental constants \citep{Kaplinghat1999}. Comprehensive reviews that motivate the search for variations of fundamental constants have been given in the literature \citep{Uzan2003, Uzan2011}. For instance, constants such as the fine structure constant, $\aEM$, or the effective electron mass, $\me$, can vary due to the introduction of non-standard electromagnetically-interacting fields \citep{Bekenstein1982}. At low redshifts, these constants have been constrained with quasar absorption lines \citep{Bonifacio2014, Kotus2017,Murphy2017} { and more recently, direct measurements from the Large Magellanic Cloud \citep{Levshakov2019}. One study has tested variations of the fine structure constant using thermonuclear supernoave \citep{Negrelli2018}}. In addition, several papers have studied the variations of these fundamental constants through their effect on the ionization history and the CMB anisotropies \citep{Battye2001, Avelino2001, Scoccola2009, Menegoni2012, Planck2015var_alp}.

In \citet[henceforth HC17,][]{Hart2017}, we provided the \planck 2015 CMB constraints on constant variations of $\aEM$ and $\me$\footnote{{The relative changes of $\me$ allows for a dimensionless rescaling of the rest mass of the electron during recombination, motivated by previous works \citep[see][for more details]{Uzan2011}.}}, in detail modeling the effects using the cosmological recombination code {\tt CosmoRec} \citep{Chluba2010b}. There we also considered explicitly time-dependent variations of $\aEM$ and $\me$ across the recombination epoch using a phenomenological power-law in redshift (see HC17), that can be motivated with \citet{Mota2004}. 
Spatial variations of the fine structure constant have also been discussed \citep{Planck2015var_alp, Smith2019}, as well as variations of the gravitational constant \citep{Galli2009b, Alvey2019}.
So far, no significant departures from the expected values of fundamental constants have been reported.

For the 2018 release of CMB anisotropy results, the \planck team was able to significantly reduce remaining systematic effects in the large-scale data \citep{Planck2018over}. Furthermore, polarised foregrounds were even more carefully subtracted in this recent analysis, leading to improved constraints from the CMB polarisation data \citep[e.g., the reionisation optical depth][]{Planck2018params}. The updated likelihood has not yet been used to constrain varying fundamental constants.

In this paper, we present the limits on the variations of fundamental constants as an update to the \planck 2015 constraints of HC17. Using the developments from the PR2 to the PR3 release of \planck data and benefiting from the reduction of systematics, especially in the $E$-mode polarisation, we derive the most stringent limits on the variations of fundamental constants from the CMB to date. These include limits on $\aEM$ and $\me$ as well as their redshift-dependence. 
The general results for $\aEM$ and $\me$ and their covariances are unaltered, however, we have expanded our discussion of the parameter degeneracies, and the interplay with the obtained Hubble parameter.
This links to the apparent Hubble tension between low- and high-redshift probes \citep{Riess2016, Bernal2016, Riess2019, Planck2015params, Planck2018params}, for which previous studies argued that variations of $\aEM$ are unable to help much \citep{Knox2019}. 
Similarly, variations of the atomic energy of hydrogen or its two-photon decay rate can only relief the tension when both extensions are included, mainly at the cost of significantly increased uncertainties but without shifting the central value by much \citep{Liu2019}.

\begin{figure*}
    \centering
    \includegraphics[width=.7\linewidth]{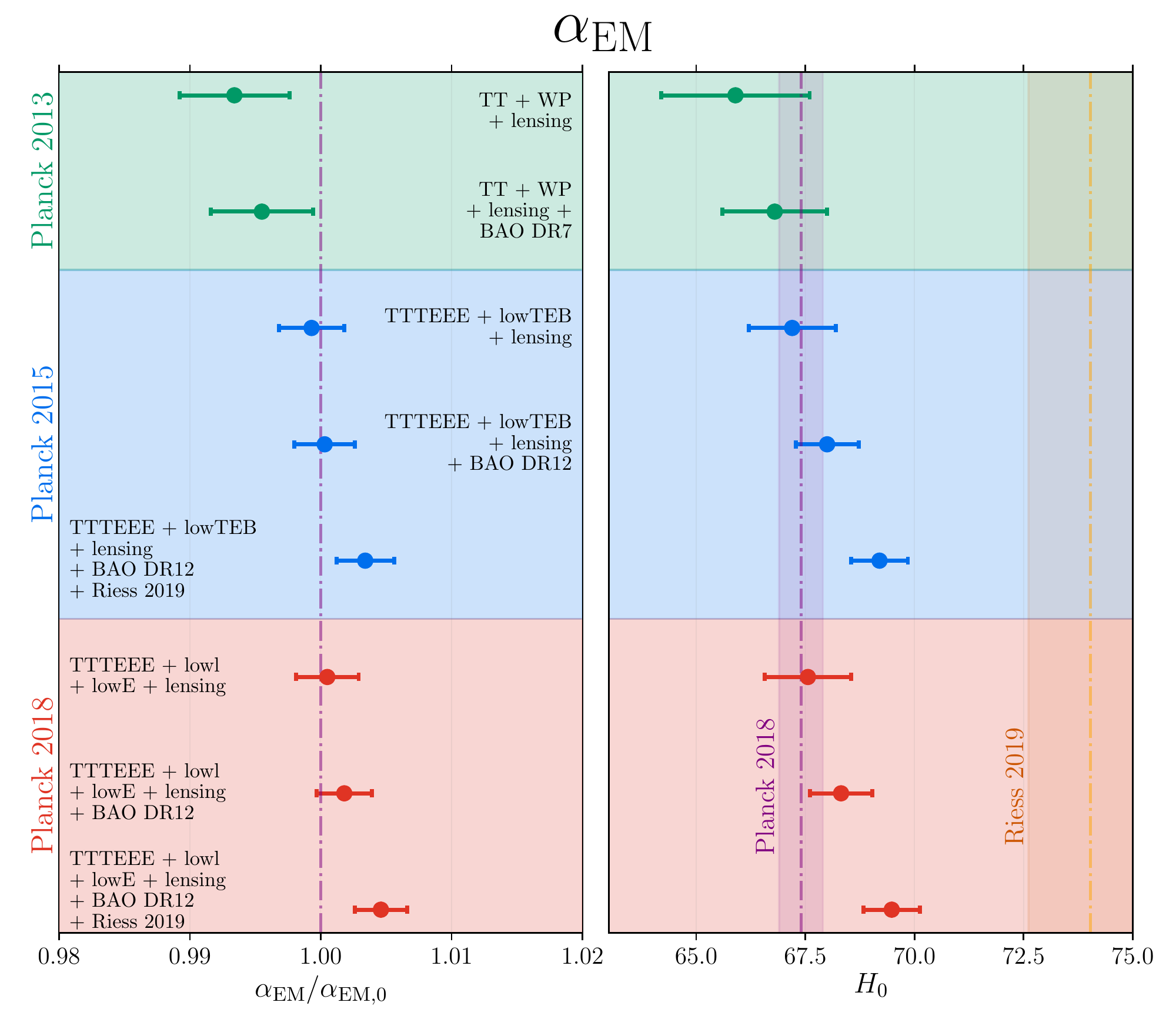}
    \centering
    \includegraphics[width=.7\linewidth]{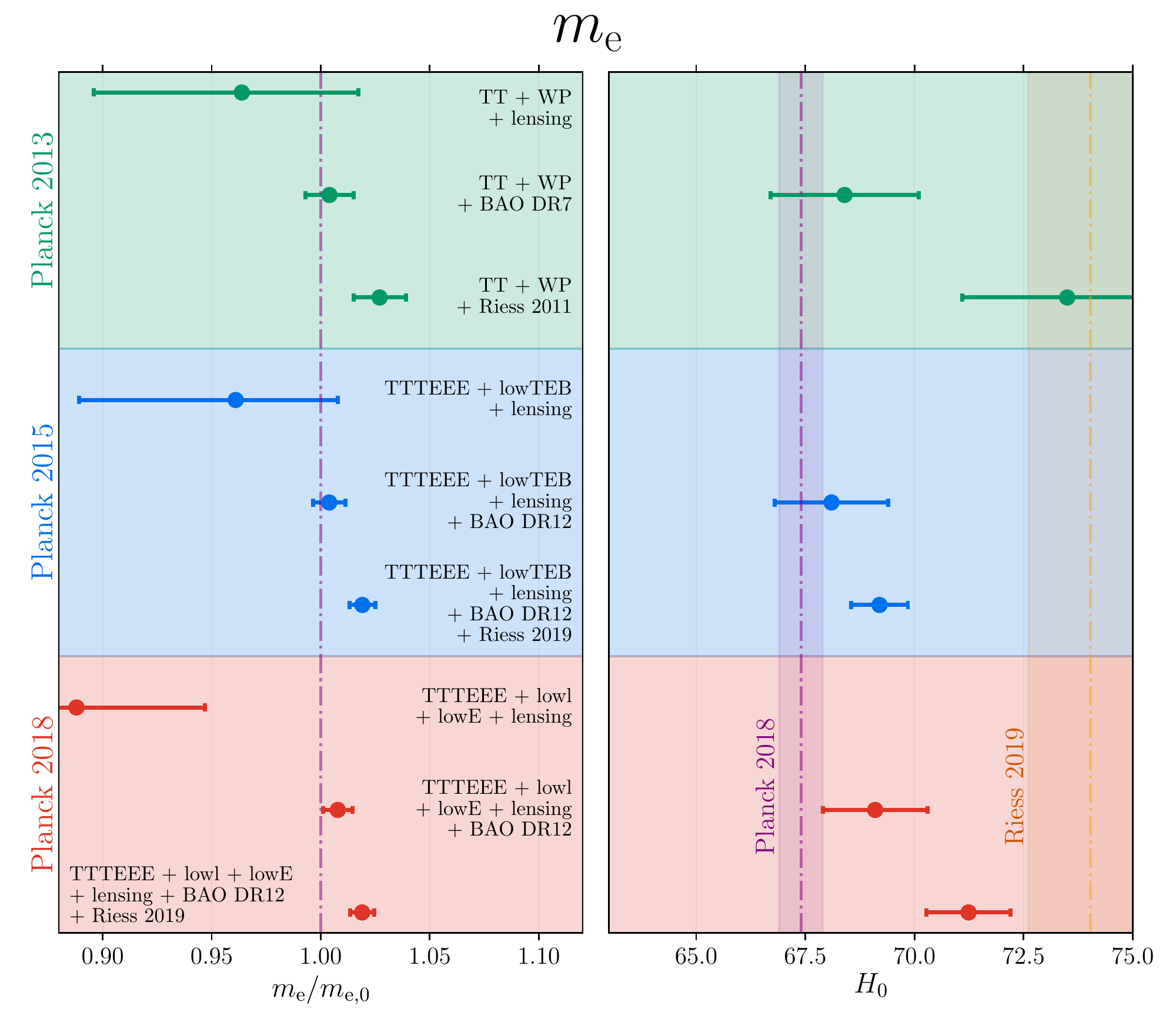}
    \caption{Constraints on the fundamental constants ({\it left}) using various combinations of \planck data included with their $\ho$ values and errors ({\it right}). The cases without {R19} data are discussed in Sect.~\ref{sec:constant}, while the addition of {R19} data is considered in Sect.~\ref{sec:tension}, when alluding to the Hubble tension.
     {\it Top:} results from the fine structure constant $\aEM$. {\it Bottom:} similar results but from the effective electron mass $\me$. Here we have redacted the data from CMB data only because the error bars are so large for $\ho$. For the $\me$ MCMC analysis, we have widened the prior on the Hubble constant such that $\ho>20$. }
    \label{fig:me_alpha}
\end{figure*}

Here, we confirm that variations of $\aEM$ indeed can only play minor role for the Hubble tension (see Fig.~\ref{fig:me_alpha}). However, a single-parameter extension that allows for variations of $\me$ indeed seems to alleviate the Hubble tension when combining various datasets at the cost of a discordant value for $\me$ during recombination and reionization\footnote{{Though we focus on recombination here, the calculation of the reionization history is the standard calculation done within {\tt CAMB}.}}. 
Indications for this behaviour were already seen for the \planck 2013 release \citep{Planck2015var_alp}, however, there the significance for the shift in $\me$ was at the $\simeq 2\sigma$ level, while here we find a discrepant value of $\me$ at $\Delta \me/\me\simeq 3.5\sigma$, thus calling for further investigation.

\vspace{-0mm}
\section{Updated constraints on varying constants} 
\label{sec:constant}
For this paper, the reference dataset is the 2018 baseline \planck data, with low-$\ell$ and high-$\ell$ data for temperature and $E$-mode polarisation power spectra, along with the lensing data from the same release \citep{Planck2018like,Planck2018Lensing}.
This is also combined with baryon acoustic oscillation \citep[BAO,][]{SDSSDR12} and supernova/Cepheid variable \citep[{R19},][]{Riess2019} data.
In this section, we will focus the discussion on \planck and BAO data only and return to the effect of adding {R19} data in Sect.~\ref{sec:tension}.

As mentioned above, the large-angle polarisation data was slightly improved in the \planck 2018 data release; however, the reference \planck dataset too has changed. In previous \planck analyses the high-$\ell$ temperature and polarisation data was combined with the low-$\ell$ temperature, $E$ and $B$-mode polarisation data, along with the CMB lensing data as well. Here the \planck reference case includes just the high-$\ell$ and low-$\ell$ temperature and $E$-mode polarisation data with CMB lensing. We do not consider this posing an issue given the lack of constraining power from the $B$-modes for these fundamental constant variations. 

Here, we examine $\aEM$ and $\me$, since these fundamental constants are the ones directly affecting the recombination process and Thomson scattering of the CMB. As illustrated in HC17, the main driving effect from a constant variation of $\aEM$ is a change in the location of the last scattering surface. This originates from the main physical change to recombination from a varying fine structure constant, which modifies atomic transition energies, changing the temperature at which the photons and baryons decouple.
Varying constants also affect specific transition rates such as Lyman-$\alpha$ and two-photon processes during recombination, though these modifications lead to far smaller changes (cf. HC17). 

The effects from $\me$ on the recombination history are very similar, even if typically $\simeq 2.5$ times smaller than for a similar variation in $\aEM$. Marked differences in the way that the Thomson cross section is changed distinguishes $\aEM$ from $\me$ variations. In particular, this effect strongly increases the geometric degeneracies between $\me$ and $\ho$ [throughout the paper in units of ${\rm km \, s^{-1} \, Mpc^{-1} }$], leading to a significantly enhanced error on $\me$ from CMB data alone (HC17).
The latter motivates a more careful consideration of the $\me$ constraints when combining CMB with external data, as we especially discuss in Sect.~\ref{sec:tension}, where we present new results for both the \planck 2015 and 2018 releases.

\begin{table*}
\centering
\begin{tabular} { l c | c c | c c}
\hline
\hline
Parameter &\planck 2018 &\planck 2018 &\planck 2018 + BAO &\planck 2018 &\planck 2018 + BAO\\
 & &  + varying $\aEM$ & + varying $\aEM$ & + varying $\me$ & + varying $\me$ \\
\hline
$\Omega_b h^2  $ &  $0.02237\pm 0.00015  $ &  $0.02236\pm 0.00015  $ &  $0.02240\pm 0.00014  $ &  $0.0199^{+0.0012}_{-0.0014}  $ &  $0.02255\pm 0.00016  $\\
$\Omega_c h^2  $ &  $0.1199\pm 0.0012  $ &  $0.1201\pm 0.0014  $ &  $0.1199\pm 0.0015  $ &  $0.1058\pm 0.0076  $ &  $0.1208\pm 0.0018  $\\
$100\theta_{MC}  $ &  $1.04088\pm 0.00031  $ &  $1.0416\pm 0.0034  $ &  $1.0436\pm 0.0030  $ &  $0.958\pm 0.045  $ &  $1.0464\pm 0.0047  $\\
$\tau  $ &  $0.0542\pm 0.0074  $ &  $0.0540\pm 0.0075  $ &  $0.0553\pm 0.0075  $ &  $0.0512\pm 0.0077  $ &  $0.0549\pm 0.0074  $\\
${\rm{ln}}(10^{10} \As)  $ &  $3.044\pm 0.014  $ &  $3.043\pm 0.015  $ &  $3.043\pm 0.015  $ &  $3.029\pm 0.017  $ &  $3.045\pm 0.014  $\\
$\ns  $ &  $0.9649\pm 0.0041  $ &  $0.9637\pm 0.0070  $ &  $0.9621\pm 0.0070  $ &  $0.9640\pm 0.0040  $ &  $0.9654\pm 0.0040  $\\
\hline
$\alpha_{\rm EM}/\alpha_{\rm EM,0}  $ & $--$   &  $1.0005\pm 0.0024  $ &  $1.0019\pm 0.0022  $ & $--$   & $--$  \\
$m_{\rm e}/m_{\rm e\,,0}  $ & $--$   & $--$   & $--$   &  $0.888\pm 0.059  $ &  $1.0078\pm 0.0067  $\\
\hline
$\ho  $ &  $67.36\pm 0.54  $ &  $67.56\pm 0.99  $ &  $68.32\pm 0.71  $ &  $46^{+9}_{-10}  $ &  $69.1\pm 1.2  $\\
\hline\hline
\end{tabular}
    \caption{Marginalised values of the fine structure constant and effective electron mass $\aEM$ and $\me$ using the \planck 2018 data along with BAO contributions. We used a wide prior for $\ho$ so that the 1$\sigma$ limit is not cut off and therefore avoids biasing the marginalised $\me$ posterior $\left(\ho>20\right)$.}
    \label{tab:alpha_me2018}
\end{table*}

\subsection{Variations of the fine structure constant: $\aEM$}
\label{sec:alp}
We find that the constraints on $\aEM$ remain largely unchanged when moving to the \planck 2018 dataset, albeit yielding slightly improved errors. 
We summarized the changes through the various \planck data releases in Fig.~\ref{fig:me_alpha}. If we compare with the CMB-only values from our previous paper, we can see a change of the fine structure constant from $\aEM/\aEMs=0.9993\pm0.0025$ to $\aEM/\aEMs=1.0005\pm0.0024$.  When we add BAO data, the value drifts slightly away from unity, $\aEM/\aEMs=1.0019\pm0.0022$. This matches the behaviour found for the 2013 and 2015 data, as seen in Fig.~\ref{fig:me_alpha}, where the fiducial value of $\aEM$ shifts to slightly higher values. The errors have not improved substantially due to the addition of higher-precision polarisation data, mainly affecting the reionization optical depth $\tau$, which is largely uncorrelated with $\aEM$. 

The values of the standard parameters when adding $\aEM$ are presented in Table~\ref{tab:alpha_me2018}. The majority of the fiducial parameters' positions and errors stay the same except for $\ns$ and $\thetaMC$. The error in $\ns$ increases by $0.7\sigma$, however, the $\theta_{\rm MC}$ error increases by an order of magnitude due to geometric degeneracies \citep[e.g.,][]{Planck2015var_alp}. The effect is not as dramatic in $\ho$, where the error is only doubled due to the non-linear contributions from some of the other parameters diluting the increase in the error. This all follows the behaviour of the constraints when $\aEM$ was added in previous works \citep{Planck2015var_alp, Hart2017}. Similarly, Table~\ref{tab:alpha_me2018} shows that when adding BAO data, the limits do not change qualitatively, except the distance ladder contributions leading to $\sigma_{\ho}=0.71$ while pulling $\ho$ sightly upwards. 

\begin{figure}
    \centering
    \includegraphics[width=0.99\linewidth]{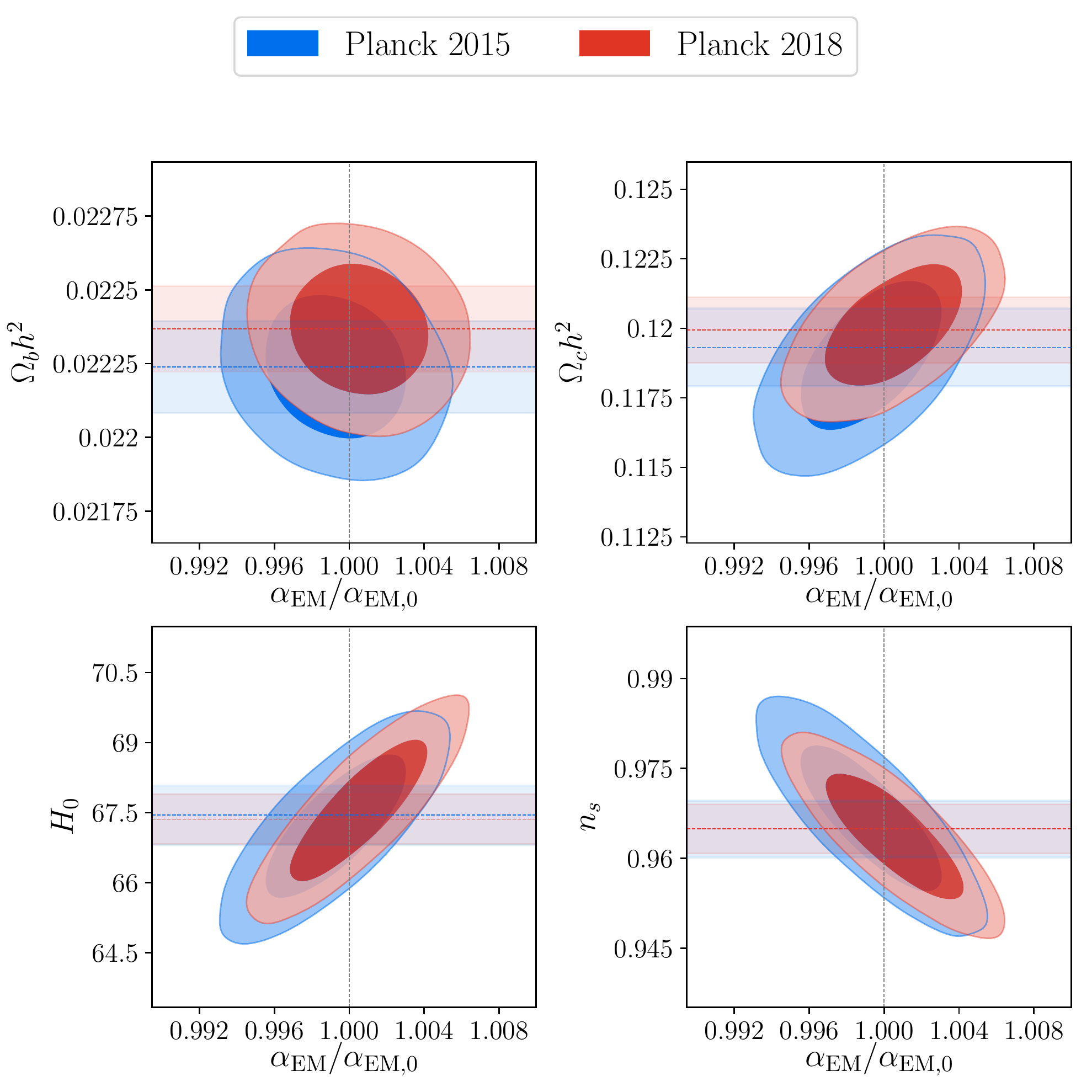}
    \caption{The posterior contours that illustrate the degeneracies between the fine structure constant $\aEM$ and the most-affected standard parameters $\{\omb,\omc,\ho,\ns\}$. 
    The $\Lambda$CDM scenario of $\aEM/\aEMs=1.0$ is added as a reference, along with the 1$\sigma$ values of \planck 2015 and 2018 standard parameters in coloured bands.}
    \label{fig:alpha2018}
\end{figure}

A selection of the most-affected standard parameter posteriors is shown in Fig.~\ref{fig:alpha2018}, where we have compared the CMB-only 2015 (red) and 2018 posteriors (blue), along with respectively coloured bands, showing the fiducial $1\sigma$ limits of these parameters. The degeneracy with $\ho$ has not changed except that the locations of the contours have shifted to slightly larger marginalised values of $\ho$, from $\ho=67.2\pm1.0$ in the 2015 release to $\ho=67.56\pm0.99$ in the 2018 release. This shift marks the main degeneracy of the $\aEM$ parameter, however, we can also clearly see a degeneracy with the tilt of the spectrum, another parameter that slightly differs between the \planck releases. Comparing the fiducial bands in Fig.~\ref{fig:alpha2018} with the locations of the contours, it is clear that the $\omb$ and $\omc$ contours have shifted in opposite directions to their 1$\sigma$ bands between the two datasets. Overall, we find consistent results for all the parameters between the two \planck releases. Addition of BAO data to the analysis does not alter the conclusions significantly.

\vspace{-3mm}
\subsection{Variations of the effective electron mass: $\me$}
\label{sec:me}
The difference in constraints for $\me$ is slightly more complicated than the picture for $\aEM$.
Although there are improvements over the \planck 2015 results, the marginalised value of $\me$ is still heavily influenced by the prior definition of $\ho$. This aspect was covered in our previous work, where we used a prior such that $\ho>40$ to conform with the initial CMB analysis with \planck, which combined \planck 2013 and {\it WMAP} data \citep{Planck2015var_alp}. However, when considering CMB-only 2018 constraints, the marginalised value of $\me$ slips further away from unity with $\ho$ departing from its fiducial CMB value. We thus used an even more extended prior $20<\ho<100$ to avoid prior-domination.

For the CMB-only results presented in Table~\ref{tab:alpha_me2018}, we applied the aforementioned wide prior; however, we have instead shown the narrow prior results in Fig.~\ref{fig:me_alpha} to better compare with previous works. Further discussion on the prior is found in Sec.~\ref{sec:malpha}, where we show the changes in the contours as the prior is adjusted. 
In 2013 and 2015, the CMB-only constraint for $\me$ were consistent with the standard value, given the large error bar due to geometric degeneracies. With the 2018 data, $\me$ drifts further below the local value, indicating a discrepancy of $\Delta \me/\me\simeq -2\sigma$ level, with extremely low value for $H_0$ (see Table~\ref{tab:alpha_me2018}). Although for the \planck 2013 and 2015 data, the value was consistent with $m_{\rm e, 0}$ to within $1\sigma$, this behaviour is not surprising given the large degeneracies between $\ho$ and $\me$ already documented in previous works on \planck data and fundamental constant variations. Adding BAO data brings $\me$ back to the standard value, restoring concordance at a slightly improved error for the 2018 data.

\begin{figure}
    \centering
    \includegraphics[width=0.99\linewidth]{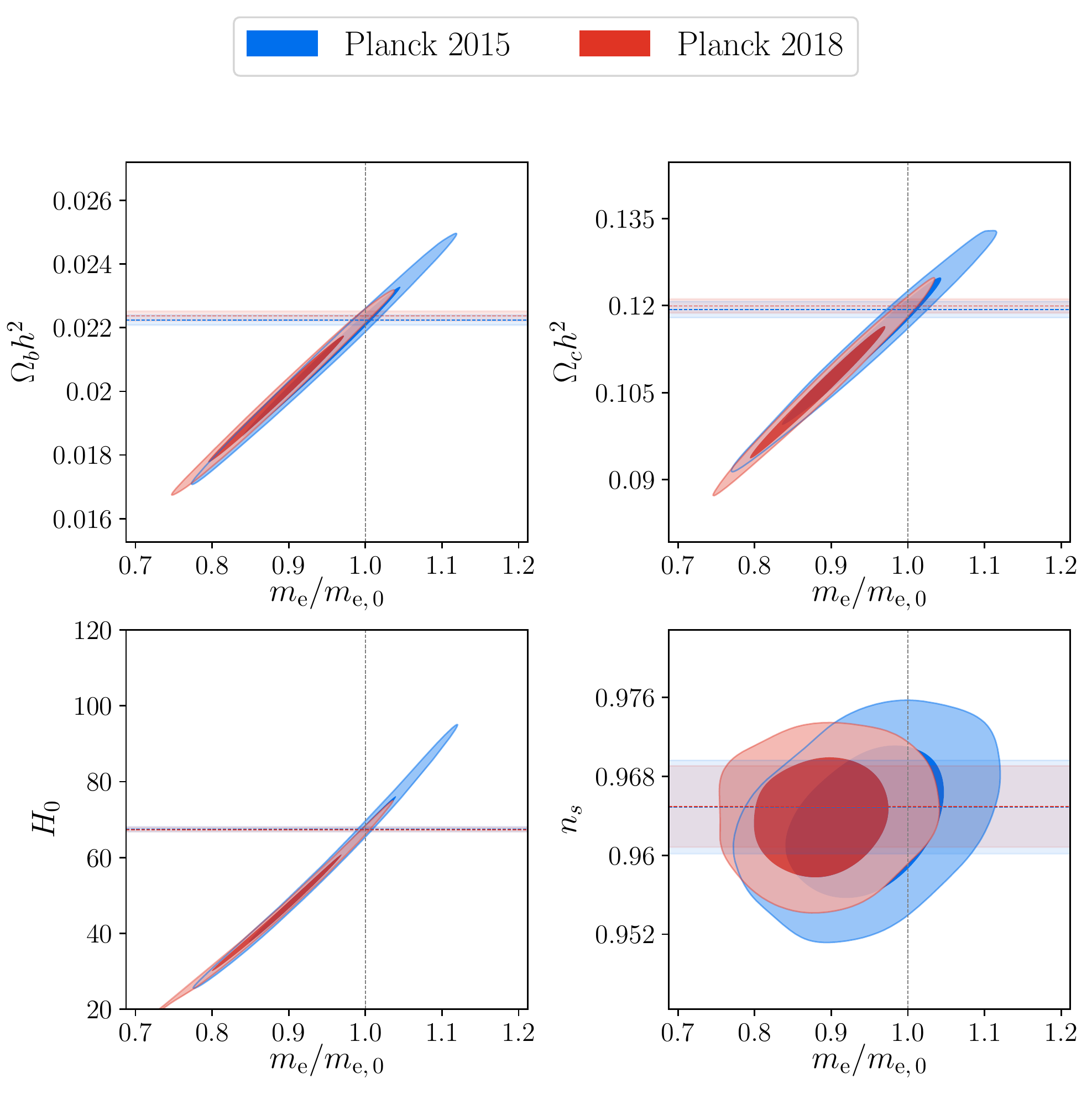}
    \caption{Same as Fig.~\ref{fig:alpha2018} except here we show the contours for $\me$. The same $\Lambda$CDM reference marker and fiducial cosmology $1\sigma$ bands have been added here. In these contours we ensured $20<\ho<100$.}
    \label{fig:me2018}
\end{figure}

While parameters like $\tau, \ns$ and $\As$ all stay within $1\sigma$ of $\Lambda$CDM when adding $\me$, we must appreciate the drifts in the baryonic and cold matter density parameters, $\omb$ and $\omc$. From Fig.~\ref{fig:me2018} and Table~\ref{tab:alpha_me2018}, we can see the sharp degeneracies of these parameters leading to $\simeq -1.8\sigma$ drifts in $\omc$ and $\simeq -2\sigma$ for $\omb$ for CMB-only constraints. These shifts are due to correlations with $\ho$ (through the well-known $\ho-\Omega_{\rm m}$ degeneracy). 
When adding BAO data, concordance with $\Lambda$CDM is again restored to within $\simeq 1\sigma$ (see Table~\ref{tab:alpha_me2018}).

\vspace{-4mm}
\subsubsection{Electron mass dependence of the Thomson cross section}
\label{sec:sigT}
As highlighted in HC17, the effect of $\me$ on the Thomson scattering cross section, $\sigT\propto\aEM^2/\me^2$, plays a distinct role for the geometric degeneracies of $\aEM$ and $\me$. {The primary effect from the rescaling of $\sigT$, outside of the recombination fraction, arises from the rescaling of the Thomson visibility function $g(z)$. This was discussed in HC17. }
For $\aEM$, the rescaling of $\sigT$ improves the obtained error at the level of $\simeq 30\%$, while for $\me$ it opens up the geometric degeneracy: if we do not include the effects on $\sigT$, the constraint changes from $\me/\mes=0.888\pm0.059$ to $\me/\mes=1.0005\pm0.0099$. In this case, we also regain a value of $\ho=67.5\pm1.7$, with an increased overall error but very close to the \planck value of $\ho=67.36\pm0.54$. The main impacts from omitting $\sigT$ come from low redshift changes in the visibility function, which give the $\me$ parameter a stronger correlation with $\ho$. 
Overall, this highlights that variations of $\me$, through the distinct effect on the Thomson optical depth open the geometric degeneracy line, an aspect that we will return to in Sect.~\ref{sec:tension}.

\begin{figure}
    \centering
    \includegraphics[width=0.99\linewidth]{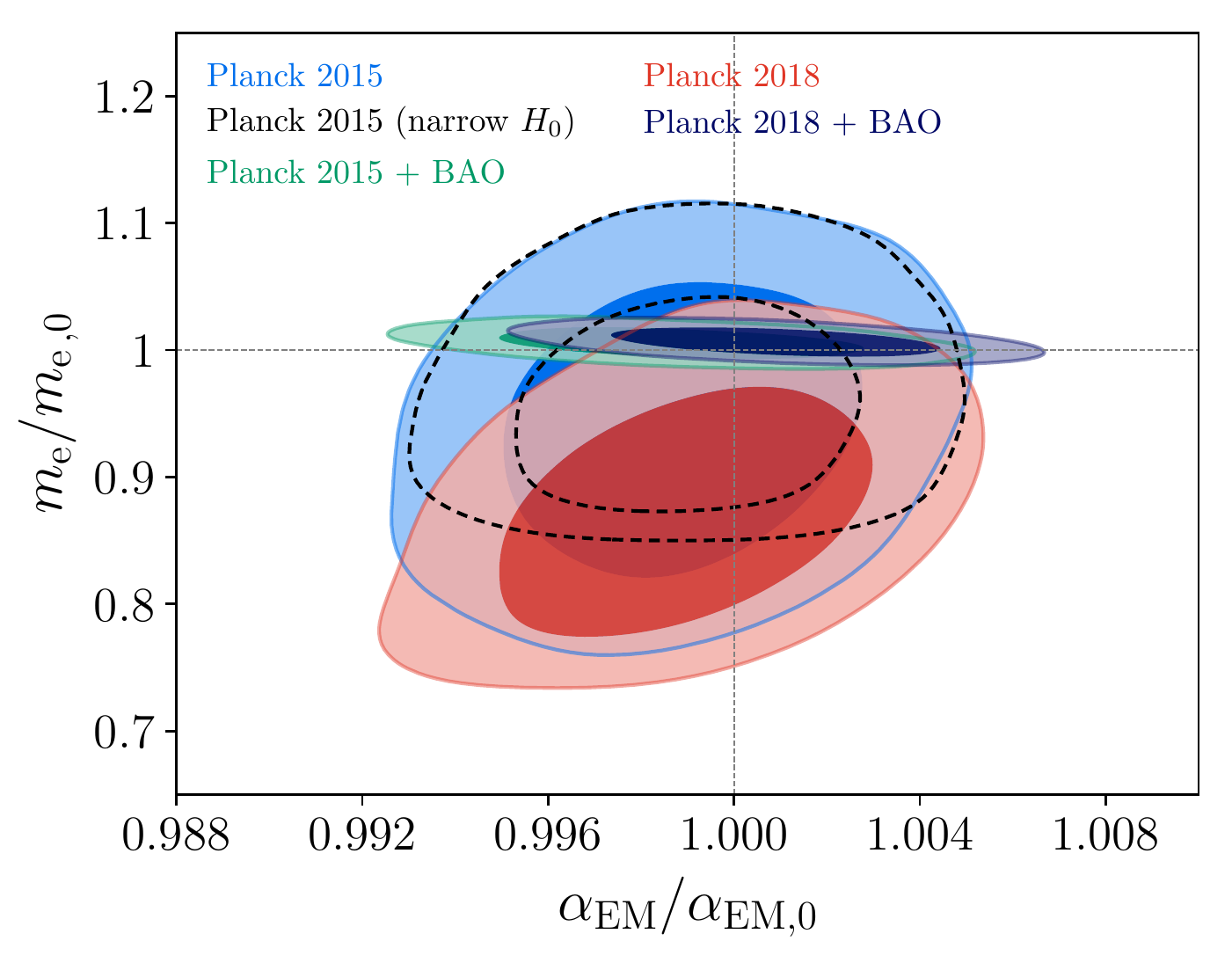}
    \caption{Posterior contours between $\aEM$ and $\me$ for the \planck 2015 and 2018 data along with BAO contributions. Note that the dashed contour shows the 2015 contour but with a tighter prior on $\ho\in \{40,100\}$.}
    \label{fig:malpha_mva}
\end{figure}

\vspace{-4mm}
\subsection{Joint variations of $\aEM$ and $\me$}
\label{sec:malpha}
In Fig.~\ref{fig:malpha_mva}, we present the contours when $\aEM$ and $\me$ are both allowed to vary. 
Here we present the posteriors for CMB-only and \planck 2018 + BAO, also comparing with the \planck 2015 results. Without BAO data, $\me$ assumes values $<m_{\rm e,0}$, covering a wide range reaching down to $\me/m_{\rm e,0}\simeq 0.75$ at $2\sigma$, as also found without extra variation of $\aEM$ (see Table~\ref{tab:alpha_me2018}). 
The two-dimensional posterior is highly non-Gaussian in this case and for the 2018 data we can even notice some truncation due to the assumed $H_0$ prior at the lower end of the contour. This is illustrated for the 2015 data run (dashed contours in Fig.~\ref{fig:malpha_mva}), mimicking the initial \planck analysis \citep{Planck2015var_alp}. {Similar to the \planck paper, we see no strong degeneracy between $\aEM$ and $\me$ and this is due to the improvements in the damping tail data which is explained in that paper.}

The introduction of BAO data tightens the constraints on $\me$ and we can also observe a small drift in $\aEM$. The obtained value changes from $\aEM/\aEMs=0.9989\pm0.0026$ for the 2015 data to $\aEM/\aEMs=1.0010\pm0.0024$ with 2018 data, both with BAO included and when simultaneously varying $\me$. By contrast, there is no drift for $\me$, with $\me/\mes=1.0056\pm0.0080$ changing to $\me/\mes=1.0054\pm0.0080$ for \planck 2015+BAO and \planck 2018+BAO, respectively.

\begin{table*}
\centering
\begin{tabular} { l  c c c c}
\hline\hline
Parameter & \planck 2018 & \planck 2018 + BAO & \planck 2018 & \planck 2018 + BAO\\
 & + varying $\aEM(z,p)$ & + varying $\aEM(z,p)$ & + varying $\me(z,p)$ & + varying $\me(z,p)$ \\
 \hline
$\Omega_b h^2  $ &  $0.02233\pm 0.00018  $ &  $0.02243\pm 0.00016  $ &  $0.0197^{+0.0012}_{-0.0015}  $ &  $0.02254\pm 0.00019  $\\
$\Omega_c h^2  $ &  $0.1198\pm 0.0017  $ &  $0.1201\pm 0.0017  $ &  $0.1045^{+0.0074}_{-0.0082}  $ &  $0.1209\pm 0.0019  $\\
$100\theta_{MC}  $ &  $1.0405\pm 0.0049  $ &  $1.0441\pm 0.0038  $ &  $0.950\pm 0.046  $ &  $1.0466\pm 0.0049  $\\
$\tau  $ &  $0.0545\pm 0.0077  $ &  $0.0549\pm 0.0076  $ &  $0.0513\pm 0.0081  $ &  $0.0544\pm 0.0075  $\\
${\rm{ln}}(10^{10} \As)  $ &  $3.044\pm 0.015  $ &  $3.043\pm 0.015  $ &  $3.030\pm 0.018  $ &  $3.044\pm 0.016  $\\
$\ns  $ &  $0.9640\pm 0.0071  $ &  $0.9622\pm 0.0069  $ &  $0.9655\pm 0.0064  $ &  $0.9645\pm 0.0066  $\\
\hline
$\alpha_{\rm EM}/\alpha_{\rm EM,0}  $ &  $0.9997\pm 0.0035  $ &  $1.0022\pm 0.0027  $ & $--$   & $--$  \\
$m_{\rm e}/m_{\rm e\,,0}  $ & $--$   & $--$   &  $0.878^{+0.057}_{-0.065}  $ &  $1.0081\pm 0.0070  $\\
$p  $ &  $-0.0011\pm 0.0035  $ &  $0.0007\pm 0.0031  $ &  $0.0014\pm 0.0043  $ &  $-0.0007\pm 0.0043  $\\
\hline
$\ho  $ &  $67.3\pm 1.4  $ &  $68.45\pm 0.88  $ &  $44^{+9}_{-10}  $ &  $69.1\pm 1.2  $\\
\hline\hline
\end{tabular}
    \caption{Marginalised parameter values for MCMC runs with \planck data along with the power law defined in \Cref{eq:power}. Both $\aEM$ and $\me$ are shown with the constrained power law parameter along with the limit when BAO is added as well.} 
    \label{tab:joint2018}
\end{table*}

\subsection{Constraints on $\aEM$ and $\me$ with time dependence}
\label{sec:timedep}
Low redshift probes of fundamental constants have shown that possible variations would have to be as small as $\sim10^{-6}$ today \citep[see the review in][for more details]{Uzan2011}. However explicitly redshift-dependent variations have been motivated by a number of theoretical models such as the string dilaton and runaway dilaton models \citep{Martins2015}. Here we consider how the fundamental constants $\mathcal{C}$ vary with redshift across the recombination epoch, using the phenomenological parametrisation 
\begin{equation}
    \mathcal{C}(z)=\mathcal{C}_0\left(\frac{1+z}{1100}\right)^p,
\label{eq:power}
\end{equation}
where $\mathcal{C}\in\{\aEM,\me\}$ for this analysis. Time-independent variations are captured by constant change to $\mathcal{C}_0$, just as in the previous sections. Varying $p$ parametrises the time-dependence. {Here our reference value reflects the \LCDM value to study the unique variations by pivotting around the maxima of the Thomson visibility function}. It leaves the position of the Thomson visibility function practically unaltered while broadening and narrowing it for negative and positive values of $p$, respectively ({Full details of the power law effects during recombination are found in \citet{Hart2017}}).

When we add time-dependent variations using the power-law model in Eq.~\eqref{eq:power} to the MCMC analyses discussed in \Cref{sec:alp,sec:me}, we obtain the results summarized in Table~\ref{tab:joint2018}.
In comparison to \planck 2015, the obtained central values and errors change marginally when $p$ is added, showing no indication for departures from $p\simeq 0$. 
Whilst $\aEM = 0.9998 \pm 0.0036$ and $p=0.0007 \pm 0.0036$ for the \planck 2015 release, the fundamental constant marginalised values are $\aEM = 0.9997\pm0.0035$ and $p=-0.0011\pm0.0035$ for the \planck 2018 data. The index of the power law $p$, is consistent with zero to within $\sim 0.3\sigma$, suggesting that this parameter is tightly constrained by the current CMB anisotropies. 
Adding BAO data, we obtain $p=0.0007\pm0.0031$, agreeing with the value for \LCDM to within $\simeq 0.2\sigma$. 
The overall posteriors remain extremely close to those presented in the \planck 2015 analysis, so that we do not repeat them here.

Turning to the case of time-varying $\me$, we again find that $p$ is consistent with zero at a fraction of a $\sigma$ (see Table~\ref{tab:joint2018}). 
Even with the large geometric degeneracies between $\ho$ and $\me$ described in Sec.~\ref{sec:me}, the only differences are small shifts of the central parameter values, which derive from the subtle differences in the effects between $\aEM$ and $\me$ on the recombination and reionization visibilities mentioned above (see Sect.~\ref{sec:sigT}). 

Our findings again highlight that explicitly time-varying $\aEM$ and $\me$ can be constrained independently of their constant variations using CMB data. A more in depth study is thus expected to lead constraints on at least two independent model parameters if the recombination physics of the $z\simeq 1100$ Universe is indeed affected.

\begin{table*}
  \centering
\begin{tabular} { l  c c c c c c}
\hline\hline
Parameter & \planck 2018 + BAO & \planck 2018 + BAO & \planck 2018 + BAO & \planck 2018 + BAO  & \planck 2018 + BAO & \planck 2018 + BAO \\
 & & + $\alpha_{\rm EM}$ & + $m_{\rm e}$ & + {R19} & + {R19} + $\alpha_{\rm EM}$ & + {R19} + $m_{\rm e}$ \\ 
\hline
$\Omega_b h^2  $ &  $0.02244\pm 0.00013  $ &  $0.02240\pm 0.00014  $ &  $0.02255\pm 0.00016  $ &  $0.02255\pm 0.00013  $ &  $0.02244\pm 0.00014  $ &  $0.02277\pm 0.00015  $\\
$\Omega_c h^2  $ &  $0.11895\pm 0.00092  $ &  $0.1199\pm 0.0015  $ &  $0.1208\pm 0.0018  $ &  $0.11791\pm 0.00090  $ &  $0.1204\pm 0.0014  $ &  $0.1229\pm 0.0017  $\\
$100\theta_{MC}  $ &  $1.04100\pm 0.00029  $ &  $1.0436\pm 0.0030  $ &  $1.0464\pm 0.0047  $ &  $1.04116\pm 0.00029  $ &  $1.0475\pm 0.0027  $ &  $1.0543\pm 0.0038  $\\
$\tau  $ &  $0.0571^{+0.0067}_{-0.0076}  $ &  $0.0553\pm 0.0075  $ &  $0.0549\pm 0.0074  $ &  $0.0602^{+0.0069}_{-0.0078}  $ &  $0.0551\pm 0.0074  $ &  $0.0533\pm 0.0074  $\\
${\rm{ln}}(10^{10} \As)  $ &  $3.048\pm 0.015  $ &  $3.043\pm 0.015  $ &  $3.045\pm 0.014  $ &  $3.052^{+0.014}_{-0.015}  $ &  $3.039\pm 0.015  $ &  $3.044\pm 0.014  $\\
$\ns  $ &  $0.9674\pm 0.0037  $ &  $0.9621\pm 0.0070  $ &  $0.9654\pm 0.0040  $ &  $0.9700\pm 0.0036  $ &  $0.9567\pm 0.0066  $ &  $0.9640\pm 0.0041  $\\
\hline
$\alpha_{\rm EM}/\alpha_{\rm EM,0}  $ & $--$   &  $1.0019\pm 0.0022  $ & $--$   & $--$   &  $1.0047\pm 0.0020  $ & $--$  \\
$m_{\rm e}/m_{\rm e\,,0}  $ & $--$   & $--$   &  $1.0078\pm 0.0067  $ & $--$   & $--$   &  $1.0190\pm 0.0055  $\\
\hline
$H_0  $ &  $67.81\pm 0.42  $ &  $68.32\pm 0.71  $ &  $69.1\pm 1.2  $ &  $68.32\pm 0.41  $ &  $69.48\pm 0.65  $ &  $71.24\pm 0.96  $\\
\hline
$\Delta\chi^2_{\rm min}$ & $--$ & $0.21$ & $-0.39$ & $--$ & $-4.71$ & $-10.92$ \\
\hline\hline
\end{tabular}
\caption{\planck 2018 marginalised results for varying $\aEM$ and $\me$ along with BAO and {R19} datasets. Reference cases for CMB+BAO and CMB+BAO+{R19} are included. We also show the change in fit, $\Delta\chi^2_{\rm min}$, for the results compared to their reference cases.}
\label{tab:HST}
\end{table*}

\section{Adding {SH0ES} data and the Hubble tension}
\label{sec:tension}
Although the degeneracy between distance measures such as $\ho$ and $\aEM$ is already evident, the geometric effects due to the scaling of the Thomson cross section creates an even larger degeneracy between $\me$ and $\ho$, as well as the baryonic and cold dark matter densities ($\omb$ and $\omc$). We should thus check if the widening error bars of $\ho$ and $\me$, as explained in Sec.~\ref{sec:me}, allow us to alleviate some of the recently discussed tensions with supernovae data \citep[e.g.][]{Riess2016, Bernal2016, Knox2019}.

In this section, we add SH0ES data (hereafter referred to as {R19}) to the analysis using a prior of $\ho = 74.03\pm1.42$ \citep{Riess2019}. Our results are summarized in Fig.~\ref{fig:me_alpha} and Table~\ref{tab:HST}. {We explicitly do not include fundamental constant variations in the low redshift data as we are simply studying the capability for these variations during recombination.}
When varying $\aEM$, a slight shift towards $\aEM/\aEMs>1$ is seen both for the \planck 2015+BAO+{R19} and \planck 2018+BAO+{R19} data. In addition, the derived value for $\ho$ moves towards the value preferred by the {R19} data. However, the overall shifts remain $\lesssim 3\sigma$ when allowing $\aEM$ to vary. This indicates that $\aEM$ variations alone cannot fully reconcile the low- and high-redshift probes of the cosmological expansion rate.

%
In contrast, when we look at the results for $\me$ with added {R19} measurements, the Hubble constant shifts to $\ho=71.24~\pm~0.96$, bringing it to within $\simeq 2\sigma$ of the {R19} value for the \planck 2018+BAO+{R19} combination. For this case, the effective electron mass is $\me/\mes=1.0190\pm0.0055$, which indicates a $\simeq~3.5\sigma$ tension with the $\Lambda$CDM value. This begs the question whether $\me$ could indeed play a role in the Hubble tension.

To understand this result a little better, we show the $\me-\aEM$ contour for the \planck 2018+BAO dataset, highlighting the corresponding value of $\ho$ in colour. As expected, the value of $\aEM$ does not to exhibit any strong preference on the value of $\ho$, while $\me$ is strongly correlated. 
%
%
As mentioned in Sect.~\ref{sec:sigT}, the distinct role of $\sigT$ opens the geometric degeneracy line when $\me$ is allowed to vary, which without rescaling $\sigT$ is not present (HC17).
When adding {R19} data, the direction towards higher values of $\ho$ opens, allowing $\me$ to settle at $\me/\mes>1$ with {R19} data is added.

\begin{figure}
    \centering
    \includegraphics[width=0.95\linewidth]{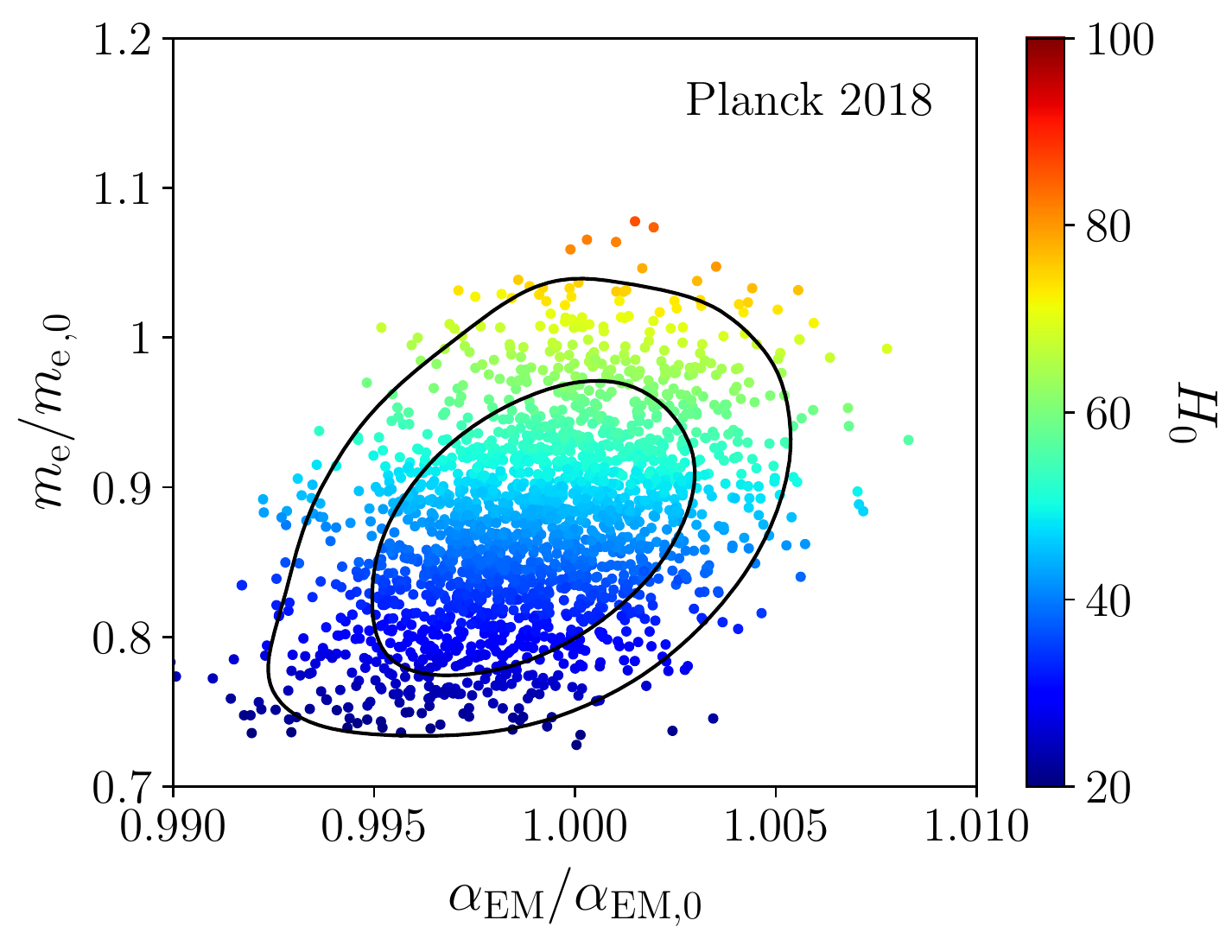}
     \caption{Two dimensional posteriors for $\aEM$ and $\me$ with the values of $\ho$ for the given samples shown in colour for \planck 2015 (upper panel) and \planck 2018 (lower panel). The prior was set to $20<\ho<100$, slightly affecting the lower end of posterior.}
    \label{fig:malpha3d}
\end{figure}

To further assess the situation, we also looked at the total $\chi^2$ from the marginalised result and compared the two cases of adding $\aEM$ and $\me$ to the reference cases of CMB + BAO and CMB + BAO + {R19}. The low positive variation of $\aEM$ coupled with the negative $\Delta\chi^2$ values shown in Table~\ref{tab:HST} indicate that the goodness-of-fit is not severely sacrificed when we add $\aEM$ and $\me$.
For the CMB + BAO case, adding $\aEM$ gives a slightly worse fit but $\me$ gives a marginally better fit. When we add {R19} data, the $\chi^2$ drops for both $\aEM$ and $\me$. We also find that the CMB-only component of the total $\chi^2$ decreases when adding $\me$ to CMB+BAO+{R19} such that $\Delta\chi^{2}_{\rm CMB}=-1.17$. This means the goodness-of-fit improves for the CMB+BAO+{R19} when $\me$ variations included.
A detailed comparison of the $\chi^2$ values is given in Table~\ref{tab:chi2}.

The tendency to allow for larger values of $\me$ and $\ho$ is even clear in the \planck 2013 data release \citep{Planck2015var_alp} when combined with the 2011 SN data \citep{Riess2011}. Though the migration of both $\me$ and $\ho$ is larger for \planck 2013 (see Fig.~\ref{fig:me_alpha}), the value of $\me$ is only $2.3\sigma$ away from \LCDM and no significant Hubble tension was yet identified back then. 
%
%
We find a similar constraint on $\me$ as for \planck 2018 when we consider a combination of \planck 2015+BAO+{R19} data, $\me/\mes=1.0191 \pm 0.0059 \,(\Delta\me/\me\simeq 3.2\sigma)$; however, the movement of $\ho$ remains more restricted, leaving a $\simeq 3.4\sigma$ tension with the {R19} data (cf. Fig.~\ref{fig:me_alpha}). This indicates that for the \planck 2018 data the geometric degeneracy line is opened more strongly when allowing $\me$ to vary.

\begin{figure}
    \centering
    \includegraphics[width=\linewidth]{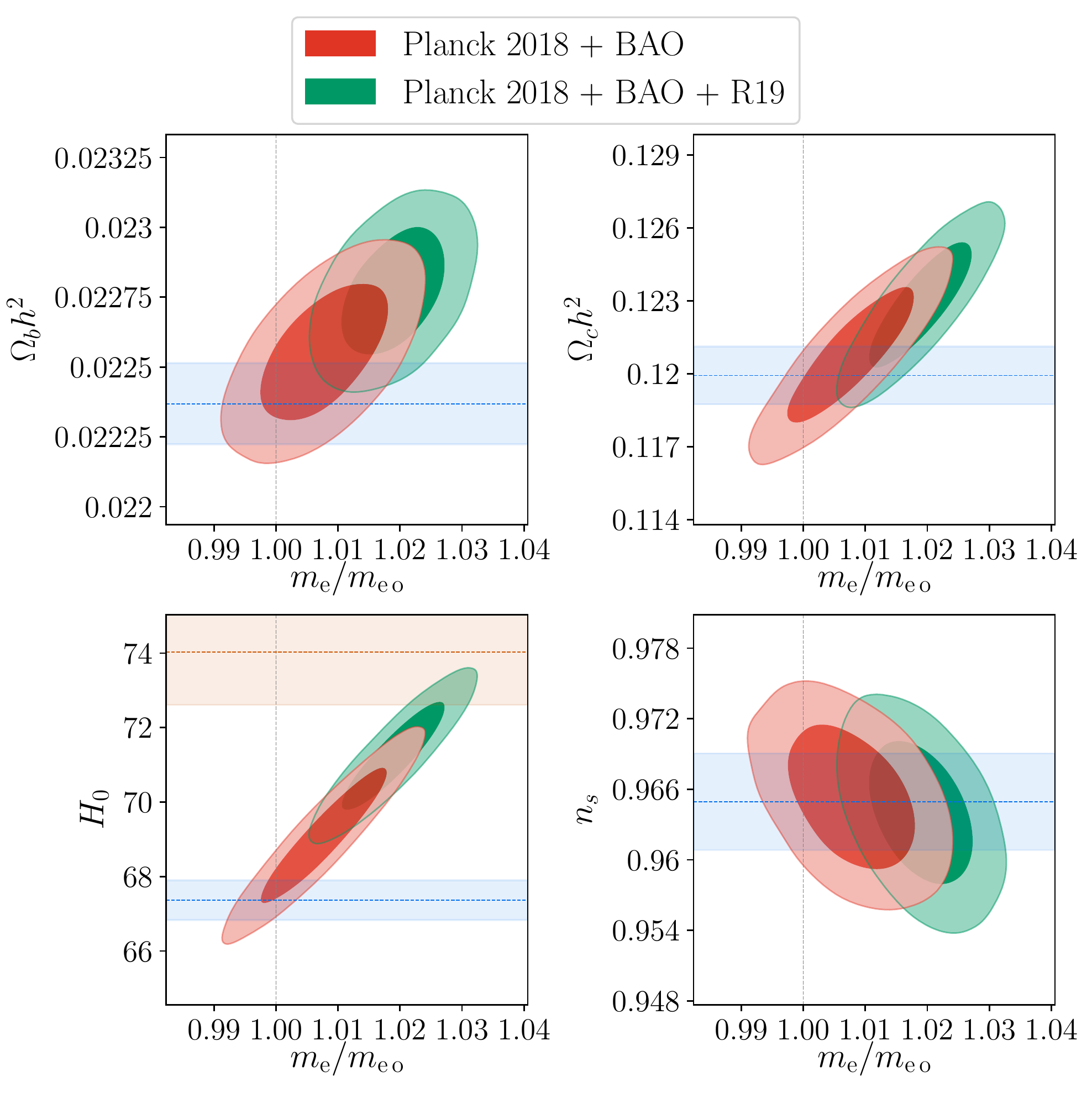}
    \caption{Marginalised contours from $\me$ with the standard parameters $\{\omb,\omc,\ho\}$. The other parameters have been omitted as they do not vary with the $\me$ changes and we have shown this in Fig.~\ref{fig:me18_BAO_full}. The blue bands represent the standard $\Lambda$CDM limits for these parameters. The orange band represents the {R19} constraint on $\ho$ \citep{Riess2019}.}
    \label{fig:me18_BAO_SN}
\end{figure}

In Fig.~\ref{fig:me18_BAO_SN}, we show the contours of some of the most affected standard parameters. A full set of parameter contours is shown in Appendix~\ref{app:results}. The contours shift away from the CMB-only posteriors (blue bands) as BAO (red) and {R19} (green) data are added. The $\ho-\me$ contour is narrower as {R19} data is added and as described above, weighted more towards the {R19} only value. We can see similar effects in $\omb$ and $\omc$, which move by $\simeq 2.5\sigma$ and $\simeq 1.8\sigma$, respectively. 
All this indicates that a discordant value of $\me$ can be traded in for an alleviation of the $\ho$ tension while affecting the standard parameters at the level of $\lesssim 2.3 \sigma$.

\begin{figure}
    \centering
    \includegraphics[width=\linewidth]{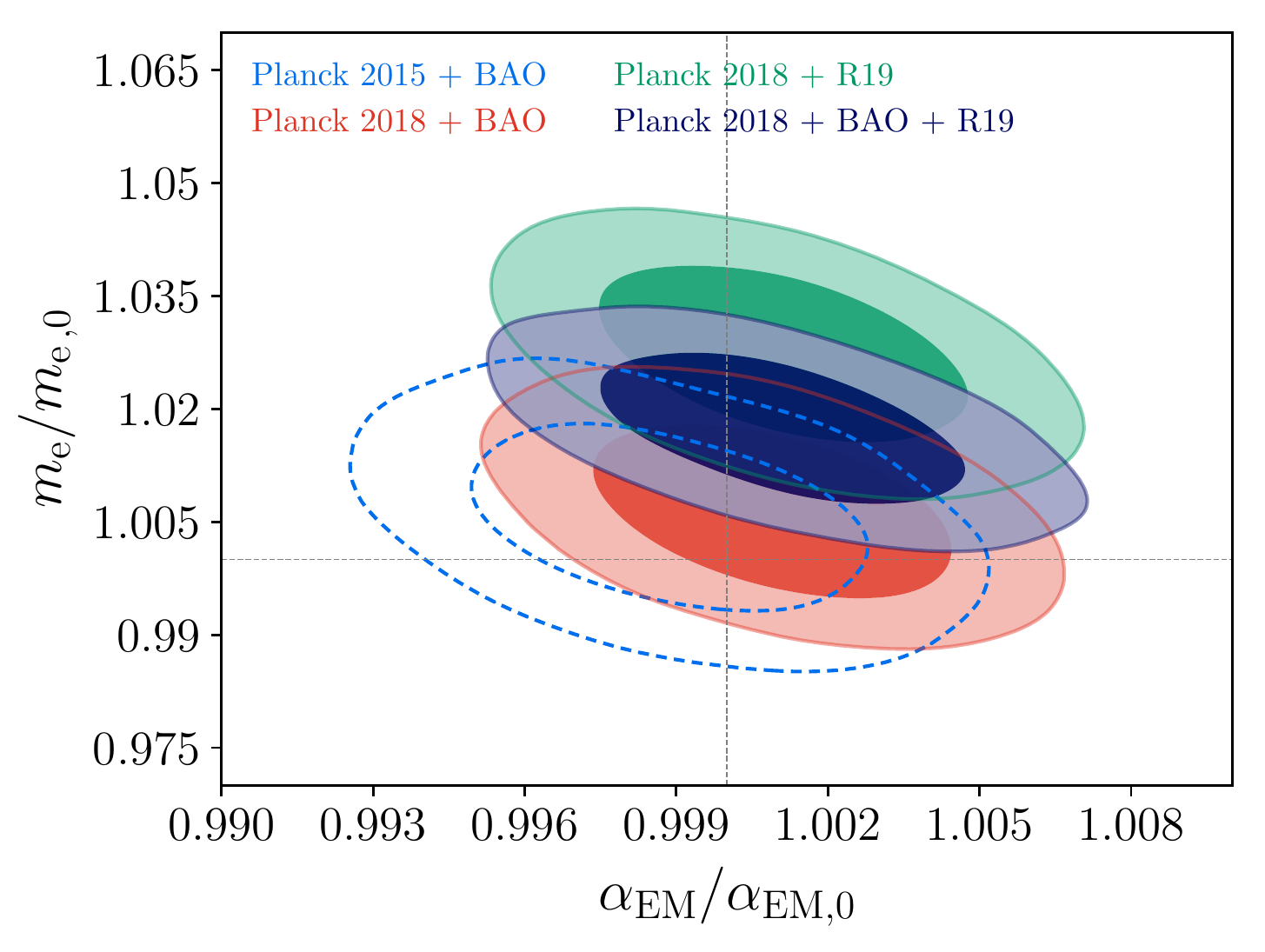}
    \caption{Probability contours between $\aEM$ and $\me$ for \planck and BAO, with added {R19} constraints as well.}
    \label{fig:malpha_mva_bao}
\end{figure}

Finally, the analysis with both $\me$ and $\aEM$ varying can be extended to include {R19} data as shown in Fig.~\ref{fig:malpha_mva_bao}. Here we have redone the analysis from Fig.~\ref{fig:malpha_mva}, however the \planck-only contours have been removed for clarity. 
Though there is a small drift in $\aEM$, the main effect from adding {R19} constraints is a migration of $\me$ away from the standard value. This is also expected from Fig.~\ref{fig:malpha3d} and further supports the perspective that variations of $\me$ could indeed play a direct role in explaining the Hubble tension.

\begin{table}
    \centering
    \begin{tabular}{l c c c}
        \hline\hline
        \planck 2018 + BAO & \LCDM & + $\Delta\aEM$ & + $\Delta\me$ \\
        \hline
        \planck (high-$\ell$+low-$\ell$) & $2789.36$ & $2790.04$ & $2789.50$ \\
        BAO & $8.61$ & $8.14$ & $8.08$ \\
        \hline
        $\Delta\chi^2$ & $--$ & $0.21$ & $-0.39$ \\
        \hline\hline
        \planck 2018 + BAO + {R19} & \LCDM & + $\Delta\aEM$ & + $\Delta\me$ \\ 
        \hline
        \planck (high-$\ell$+low-$\ell$) & $2791.65$ & $2792.65$ & $2790.48$ \\
        BAO & $7.59$ & $7.65$ & $9.77$ \\
        {R19} & $16.25$ & $10.48$ & $4.31$ \\
        \hline
        $\Delta\chi^2$ & $--$ & $-4.71$ & $-10.92$ \\
        \hline\hline
    \end{tabular}
    \caption{Changes in the goodness-of-fit $\chi^2$ when $\aEM$ or $\me$ is added to the \LCDM model. Here the \LCDM cases are references to compare the $\Delta\chi^2$ from each of the added fundamental constant variations. All the quoted $\chi^2$ values are the fits from the final marginalised results quoted in Table~\ref{tab:HST}.}
    \label{tab:chi2}
\end{table}

\section{Conclusion}
We provided updated constraints on the variation of the fundamental constants $\aEM$ and $\me$, closely following the discussion of HC17 for \planck 2015 data.
When omitting {Riess 2019} data, we find no significant difference between the results of the 2015 and the present analysis for variations of $\aEM$ and $\me$ (see Fig.~\ref{fig:me_alpha}).
As expected, the addition of improved polarisation information from \planck 2018 leads to slightly improved errors on $\aEM$ and $\me$ with shifts between the values from the 2015 and 2018 data combinations remaining below $\simeq 1\sigma$ (see Table~\ref{tab:alpha_me2018} and \ref{tab:joint2018} for summary of the parameter values).

In addition to the update of the \planck 2015 analyses, we also extended the discussion to combinations including {R19} data (Sect.~\ref{sec:tension}).
{R19} data pulls both $\aEM$ and $\me$ above their standard values; however, only for the combination \planck 2018+BAO+{R19} does the migration exceed the $3\sigma$ threshold (see Fig.~\ref{fig:me_alpha} and Table~\ref{tab:HST}). Simultaneously, we find the value for $\ho$ to move closer to that obtained from {R19} data.
Improvements in the $\chi^2$-values further indicate that variations of $\me$ could indeed play a role in the low- versus high-redshift Hubble tension.

As already alluded to in HC17, the distinct role of $\me$ in the value of $\sigT\simeq \aEM^2/\me^2$ opens the geometric degeneracy line to enable {R19} data to overcome the tight grip of \planck data on $\ho$. When neglecting the rescaling of $\sigT$ with $\me$ this degree of freedom closes. 
Our analysis thus suggests that a delicate interplay between the low- and high-redshift ionization history could indeed influence our interpretation of the cosmological datasets. A shift of $\me$ away from the standard value appears to enable this. More general alternatives could be a modified recombination history and simultaneously altered reionization history. Future works should thus investigate a possible integration of the reionization and recombination processes to more accurately model fundamental constant variations across cosmic time. 

Models with explicit time-dependence of $\aEM$ and $\me$ should furthermore be more carefully studied. As pointed out in \citet{Poulin2019}, models of early-dark energy could also play a role in the Hubble tension. Similar physical mechanisms could give rise to varying constants, potentially linking the effects to the same underlying scalar field. 
Recently, models with positive spatial curvature have too enriched the discussion on the Hubble tension \citep{DiValentino2019}. Including fundamental constant variations may alleviate the discrepancies in lensing in a similar way to that of a closed Universe. Also, spatial variations of fundamental constants could be present, potentially linking the CMB anomalies seen at large-angular scales \citep{Planck2019anomalies, Smith2019}.
We look forward to exploring these possibilities in the future.

Finally, we stress that there seems to be a marked difference between the \planck 2015 and 2018 data. For \planck 2015, we find similar constraints on $\me$ but the shift in the Hubble parameter is more mild and unable to reconcile $\ho$ (see Fig.~\ref{fig:me_alpha}). This indicates that improvements to the \planck 2018 polarization data opened the aforementioned geometric degeneracy more strongly. This calls for further investigations of systematic effects to clearly identify the origin of our findings regarding the Hubble tension.

{\small  
\section*{Acknowledgments}
We would like to thank Richard Battye and Aditya Rotti for helpful discussions. We would also like to thank Antony Lewis for his continued support whilst using the {\tt CosmoMC} MCMC code and Eleonora Di Valentino for advice on the use of the \planck 2018 likelihood code. We would like to thank Silvia Galli for useful discussions surrounding fundamental constants and the Planck likelihood. {We thank Vivian Poulin for interesting discussions surrounding the Hubble tension and useful extensions to the analysis. We would like to thank the reviewer for their helpful feedback regarding fundamental constant models and transparency.}
LH was funded by the Royal Society through grant RG140523.
JC was supported by the Royal Society as a Royal Society University Research Fellow at the University of Manchester, UK.
This work was also supported by the ERC Consolidator Grant {\it CMBSPEC} (No.~725456) as part of the European Union's Horizon 2020 research and innovation program.

\bibliographystyle{mn2e}
\bibliography{Lit}
}

\appendix
\section{Extra marginalised results}
\label{app:results}
\begin{figure*}
    \centering
    \includegraphics[width=.9\linewidth]{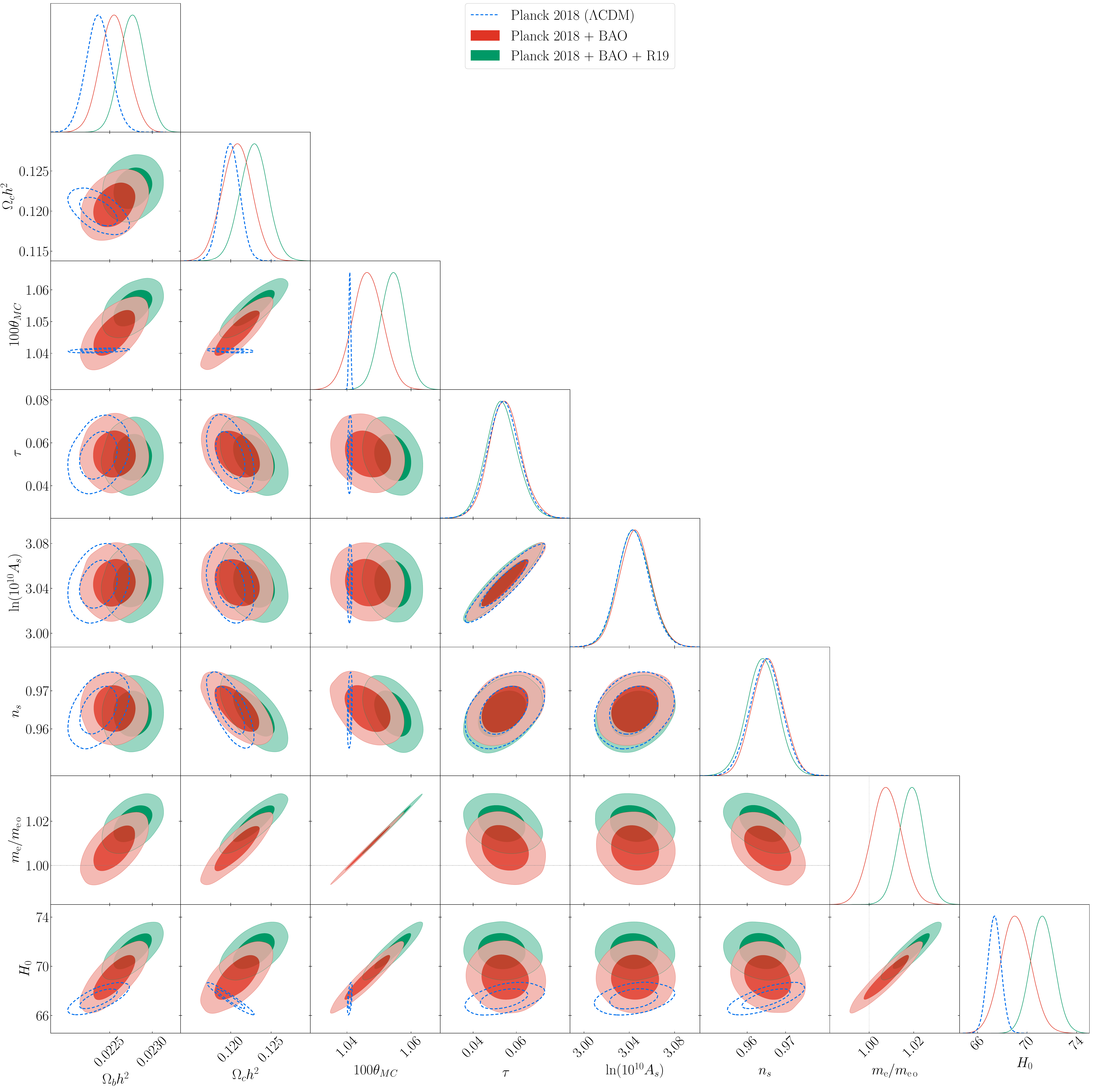}
    \caption{Fully marginalised results from \planck + BAO and \planck + BAO + {R19} and variations of $\me$ with CMB only as a reference (blue-dashed). This figure includes contributions from $\{\tau,\ns,\logA\}$ which are effectively decorrelated from $\me$. The \LCDM value of $\me=1$ is added as a dashed line.}
    \label{fig:me18_BAO_full}
\end{figure*}

For transparency, in Fig.~\ref{fig:me18_BAO_full}, we show the full parameter constraints for \LCDM parameters and $\me$ when BAO and {R19} data is added. 

\label{lastpage}
\end{document}

%% file: Befehle.tex
\newcommand{\beq}{\begin{equation}}   %

\newcommand{\eeq}{\end{equation}}   %

\newcommand{\beqa}{\begin{eqnarray}}   %

\newcommand{\eeqa}{\end{eqnarray}}   %

\newcommand{\beal}{\begin{align}}
\newcommand{\enal}{\end{align}}

\newcommand{\bspl}{\begin{split}}

\newcommand{\espl}{\end{split}}

\newcommand{\bsub}{\begin{subequations}}

\newcommand{\esub}{\end{subequations}}

\newcommand{\bmulti}{\begin{multline}}   %

\newcommand{\beqm}{\begin{mathletters}}   %

\newcommand{\eeqm}{\end{mathletters}}   %

\newcommand{\me}{m_{\rm e}}

\newcommand{\sigT}{\sigma_{\rm T}}

